\PassOptionsToPackage{table,dvipsnames}{xcolor}

\documentclass[sn-vancouver,square,numbers]{sn-jnl}

\usepackage{framed}
\usepackage[lofdepth,lotdepth]{subfig}
\usepackage{colortbl}
\usepackage{wrapfig}
\usepackage{float}
\usepackage{multirow, makecell}
\usepackage{rotating}
\usepackage{amsmath}
\usepackage{breqn}
\usepackage{amsbsy}
\usepackage{booktabs}
\usepackage{setspace}
\usepackage{amssymb}
\usepackage[section]{placeins}

\newcommand\Rey{\mbox{\textit{Re}}}  

\usepackage{ulem}

\theoremstyle{thmstyleone}%
\theoremstyle{thmstyletwo}%

\theoremstyle{thmstylethree}%

\jyear{2022}
\begin{document}

\title[Effects of wavelength on vortex structure and TKE]{Effects of wavelength on vortex structure and turbulence kinetic energy transfer of flow over undulated cylinders}


\author[1]{\fnm{Kathleen} \sur{Lyons}}\email{lyons6@wisc.edu}
\author[2]{\fnm{Raúl Bayoán} \sur{Cal}}\email{cal@me.pdx.edu}
\author*[1]{\fnm{Jennifer A.} \sur{Franck}}\email{jafranck@wisc.edu}

\affil*[1]{\orgdiv{Department of Engineering Physics}, \orgname{University of Wisconsin--Madison}, \orgaddress{\city{Madison}, \state{WI}, \country{United States}}}

\affil[2]{\orgdiv{Department of Mechanical \& Materials Engineering}, \orgname{Portland State University}, \orgaddress{\city{Portland}, \state{OR}, \country{United States}}}

\abstract{
Passive flow control research is commonly utilized to provide desirable drag and oscillating lift reduction across a range of engineering applications. This research explores the spanwise undulated cylinder inspired by seal whiskers, shown to reduce lift and drag forces when compared to smooth cylinders. Although the fluid flow over this unique complex geometry has been documented experimentally and computationally, investigations surrounding geometric modifications to the undulation topography have been limited, and fluid mechanisms by which force reduction is induced have not been fully examined. Five undulation wavelength variations of the undulated cylinder model are simulated at Reynolds number $\Rey=250$ and compared with results from a smooth elliptical cylinder. Vortex structures and turbulence kinetic energy (TKE) transfer in the wake are analyzed to explain how undulation wavelength affects force reduction. Modifications to the undulation wavelength generate a variety of flow patterns including alternating vortex rollers and hairpin vortices. Maximum force reduction is observed at wavelengths that are large enough to allow hairpin vortices to develop without intersecting each other and small enough to prevent the generation of additional alternating flow structures. The differences in flow structures modify the magnitude and location of TKE production and dissipation due to changes in mean and fluctuating strain. Decreased TKE production and increased dissipation in the near wake result in overall lower TKE and reduced body forces. Understanding the flow physics linking geometry to force reduction will guide appropriate parameter selection in bio-inspired design applications.
}

\maketitle

\section{Introduction}\label{sec:intro}
Reduction of drag and oscillating lift forces from flow over a bluff body is desirable in many engineering applications to conserve energy, reduce material costs, and lower fatigue-induced stresses on a structure. Towards this goal, various passive flow control methods have been implemented such as the addition of surface roughness, ridges, and helical strakes to an otherwise smooth surface \cite{zhang_numerical_2016,choi_control_2008}, in many instances designed for a specific application. Another potential solution is the bio-inspired design based on the unique undulated geometry of seal whiskers, which is shown to dramatically reduce drag and oscillating lift forces when compared to a smooth cylinder \cite{hanke_harbor_2010,witte_wake_2012,hans_whisker-like_2013}. This research focuses on the vortex structures and turbulent mechanisms within the wake of the seal whisker-inspired undulated cylinder and examines the influence of undulation wavelength on force reduction.

Understanding the flow over a smooth circular cylinder and geometric modifications continues to be important \cite{townsend_momentum_1949,bloor_measurements_1966,roshko_perspectives_1993,williamson_vortex_1996,yan_study_2021,mao_spanwise_2021,li_numerical_2020}.
Disruptions to the geometry along the spanwise direction can break up the nominally two-dimensional wake structures that are primarily responsible for large force fluctuations \cite{choi_control_2008}. In their computational work, Zhang et al. \cite{zhang_numerical_2016} compared several shape-modified circular cylinders with passive flow control including ridged, O-ringed, linear wavy, and sinusoidal wavy cylinders at $\Rey=5000$. Only the linear wavy and sinusoidal wavy cylinders appreciably reduced lift and drag forces, and they hypothesized that the wavy cylinder topography modified the free shear layers and stabilized the wake.

The wavy cylinder geometry is defined by a diameter that varies sinusoidally between a maximum and minimum value along the spanwise direction at a specific wavelength. Using dye visualization and pressure measurements, Ahmed \& Bays-Muchmore \cite{ahmed_transverse_1992} demonstrated three-dimensional boundary layer separation lines and the roll up of the boundary layer into streamwise vortices near locations of maximum diameter. They measured a reduction in mean drag for four wavy cylinder models as compared with a smooth circular cylinder. Zhang et al. \cite{zhang_piv_2005} found that the wavy cylinder geometry produced overall lower turbulence kinetic energy (TKE) levels in the near wake which contributed to drag reduction, however, individual transport terms were not examined. Lam \& Lin \cite{lam_effects_2009} investigated the flow over the wavy cylinder geometry for a range of wavelength and amplitude variations demonstrating a nonlinear trend of forces with respect to wavelength and two locations of minimum drag. They classified the flow into three flow regimes with respect to wavelength using spanwise vorticity to characterize three-dimensional flow structure distortion and vortex formation length variation. Similar to Ahmed \& Bays-Muchmore, at wavelengths where forces were minimal, they hypothesized that the introduction of additional streamwise vorticity tended to stabilize the two-dimensional spanwise vorticity of the free shear layers and thus prevent roll-up \cite{lam_effects_2009}.

Another common three-dimensional modification is the addition of helical strakes, which has been shown to reduce vortex shedding coherence and prevent frequency lock-in \cite{zhou_study_2011}. An extension of this variation is the helically twisted elliptical cylinder formed by rotating an elliptical cross-section along the spanwise direction. This geometry has been shown to reduce drag when compared to smooth and wavy cylinders \cite{jung_large_2014,kim_flow_2016} and displays a nonlinear drag versus wavelength trend similar to that seen by Lam \& Lin for wavy cylinders \cite{kim_flow_2016}. Kim et al. \cite{kim_flow_2016} noted minimal drag and oscillating lift forces were seen at wavelengths where vortex shedding is suppressed.

The modified oscillatory response is likely responsible for the ability of seals to track prey via hydrodynamic trail following \cite{dehnhardt_seal_1998,dehnhardt2001,schulte-pelkum_tracking_2007}. The whiskers are sensitive to disturbances in the water as they have been shown to minimize vortex-induced-vibration (VIV), resulting in a larger signal to noise ratio \cite{hanke_harbor_2010,miersch_flow_2011,kottapalli_harbor_2015}. Hanke et al. \cite{hanke_harbor_2010} showed that flow over the harbor seal whisker geometry resulted in a 40\% reduction in average drag and 90\% reduction in oscillating lift forces compared with a smooth cylinder at $\Rey=500$. Given these qualities, the undulated seal whisker geometry has inspired research of biomimetic designs such as flow sensors \cite{beem_calibration_2013,kottapalli_harbor_2015} and turbine blades \cite{shyam_application_2015,ahlman_wake_2020} among others.

However, the seal whisker geometry has greater complexity than the previously mentioned passive control surfaces due to out-of-phase surface undulations along both the streamwise and transverse directions. Experiments employing real seal whiskers have shown vibrations over a broad range of frequencies when subjected to disturbances in the incoming flow \cite{Murphy2017}. PIV in the wake of real elephant seal whiskers displayed faster wake recovery for flow over the undulating elephant seal whisker versus a smooth whisker \cite{bunjevac_wake_2018}. The specific geometry and parameters generated by Hanke et al. are widely used in fluid flow investigations, with only a handful of papers exploring geometry modifications, or examining how changing the undulation parameters affects the forces and/or resulting wake structures. Simulations by Witte et al. \cite{witte_wake_2012} compared the undulated topography with two different wavelengths found no change in drag but considerable reduction in root-mean-square (RMS) lift for the higher wavelength model. Hans et al. \cite{hans_whisker-like_2013} simulated flow over the topography with and without undulations and concluded that optimal force reduction was achieved with the inclusion of undulations along both the thickness and chord length. A 90-degree offset between the pair of undulations produced the required secondary vortex structures to stabilize the shear layer as reported by Liu et al. \cite{liu_phase-difference_2019} and further supported by Yoon et al. \cite{yoon_effect_2020}. 

Due to the complexity of the geometry, previous investigations have primarily focused on demonstration of drag and oscillating lift reduction and addressed the impact of a limited number of geometry modifications. While Hans et al. \cite{hans_whisker-like_2013} simulated four models, Yoon et al. \cite{yoon_effect_2020} simulated seven. Both investigations demonstrated the importance of undulations in the chord and thickness directions. Liu et al. simulated a larger collection of models by modifying both the amplitude and wavelength, however, the two undulation amplitude values were dependent on one another. Work by Lyons et al. \cite{lyons_flow_2020} was the first to systematically investigate each of the geometric parameters by redefining them independent from one another and simulating 16 modified feature combinations. Lyons et al. identified the most identified the most important geometric parameters affecting the flow response as the aspect ratio, the two undulation amplitudes, and the undulation wavelength. The significant importance of the aspect ratio is expected as the effect of modifying the thickness-to-chord ratio of an ellipse has been extensively studied \cite{lindsey_drag_1938,delany_low-speed_1953,faruquee_effects_2007}. The interplay between the two undulation amplitudes of the whisker-inspired geometry and their modification was described by Yuasa et al. \cite{yuasa_simulations_2022}, however wavelength has yet to be thoroughly investigated.

This paper varies undulation wavelength utilizing detailed computations and assesses the impact on the forces and wake structures through a TKE analysis. The application of TKE analysis has been used successfully to provide a comparison between experimental results and turbulent theory \cite{townsend_momentum_1949}, explore improvements in turbulence modelling \cite{liu_measurement_2004,tian_new_2020}, and clarify underlying physics of complex flow structures \cite{schanderl_structure_2017}. 
While the correlation between lower TKE in the near wake and reduced forces on the body has been well established for bluff body flows \cite{yoon_effect_2020,chu_three-dimensional_2021,lin_effects_2016,zhang_piv_2005}, the fundamental processes by which the various geometries lead to lower TKE (and hence force reduction) are not well understood. This work investigates the flow physics to provide a link between geometric topography, specifically undulation wavelength, and force reduction through detailed simulations of the undulated cylinder at various wavelengths. A deeper understanding of how geometric effects influence this complex flow will progress the development of whisker-inspired design for passive flow control and biomimetic sensing applications. 

Section \ref{sec:methods} introduces the computational methods and the whisker-inspired model. The force results, instantaneous flow structures, and turbulence terms are compared and discussed in Section \ref{sec:results}, and the summary and conclusions are presented in Section \ref{sec:conclusion}.

\section{Numerical Methods}\label{sec:methods}
\subsection{Flow Simulation Details}
Direct numerical simulations (DNS) of the flow are performed using the open-source software \textit{OpenFOAM} \cite{weller_tensorial_1998}. The governing equations are the incompressible Navier-Stokes and continuity equations,
\begin{align}
    \label{e:NSE}
    \frac{\partial {u}_i}{\partial t} + \frac{\partial {u}_i {u}_j}{\partial x_j} &= -\frac{1}{\rho}
    \frac{\partial {p}}{\partial x_i} + \nu \frac{\partial^2 {u}_i}{\partial x_j \partial x_j} \\[1.5ex]
    \label{e:con}
    \frac{\partial {u}_j}{ \partial x_j} &= 0 \, \text{,}
\end{align}
\noindent where $p$ is pressure, $\nu$ is kinematic viscosity, $\rho$ is density, and $u_j$ represents each of the three instantaneous velocity components. The \textit{OpenFOAM} libraries implement a second-order accurate finite-volume scheme with Gaussian integration and linear cell center-to-cell face interpolation. The matrix equations are solved using a generalized geometric-algebraic multi-grid method with Gauss-Seidel smoothing. The pressure-implicit split-operator (PISO) algorithm is used for pressure-velocity coupling, and time stepping is completed using a second-order accurate backward scheme. The timestep size is allowed to vary while maintaining a Courant–Friedrichs–Lewy number less than one, with an average timestep of 0.02 convective time units.

\begin{figure}
    \centerline{\includegraphics[width=4in]{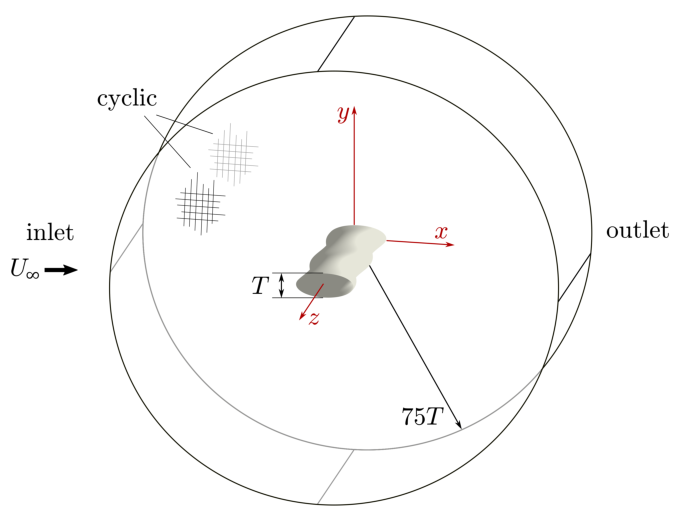}}
    \caption{Computational domain with a radius of $75T$ and a two wavelength span. Model is oriented at zero angle-of-attack and uniform velocity is in the positive $x$-direction.}
    \label{fig:domain}
\end{figure}

The computational domain is sketched in figure \ref{fig:domain} where uniform freestream velocity $U_{\infty}$ in the $x$-direction and zero pressure gradient conditions are applied at the inlet, and no-slip conditions are enforced on the model wall. At the outlet boundary, fixed pressure and zero velocity gradient conditions are set. The domain has a radius of $75T$, where $T$ is the average thickness, and two wavelengths are modelled for each geometry in the spanwise direction which is shown to be sufficient for resolving three-dimensional effects in the wake of seal whisker topographies \cite{kim_effect_2017}. Spanwise periodicity is enforced with cyclic boundary conditions that are applied to the front and back $x$-$y$ planes of the domain, thus, tip effects are not modelled. Each model is oriented at zero angle-of-attack with respect to the chord length such that the average lift coefficient is negligible. Simulations are completed at $\Rey=250$ where $\Rey$ is based on average thickness, $\Rey=U_{\infty} T/\nu$. The choice of $\Rey$ is driven by biological relevance, motivated by real-life seal foraging speed and whisker thickness \cite{lesage_functional_1999}. Furthermore, the low $\Rey$ enables the use of DNS for flow simulation with a highly resolved mesh and clear visualization of shed flow structures. It is noted that the flow over the smooth ellipse remains in a nominally laminar periodic vortex shedding regime. The introduction of three-dimensional instabilities during the transition to turbulence for flow over a smooth circular cylinder may begin at $\Rey<200$ \cite{williamson_vortex_1996,behara_wake_2010}, but the elliptical shape works to move the onset of instabilities to a higher $\Rey$.

\subsection{Calculation of Turbulence Kinetic Energy}
To understand the flow mechanisms responsible for the variations between models, an analysis of the turbulence kinetic energy budget is performed. To begin the analysis, the Reynolds decomposition
\begin{equation}
    \boldsymbol{u}(x,y,z,t) = \overline{\boldsymbol{U}}(x,y,z) + \boldsymbol{u}'(x,y,z,t)
\end{equation}
where overline represents a time-averaged quantity and prime represents a fluctuating quantity, is applied to the momentum equation to derive the Reynolds stress equation \cite{pope_turbulent_2000}. Due to the geometric variation across span, no averaging is done across the spatial directions. Half of the trace of the Reynolds stress tensor represents the turbulence kinetic energy
\begin{equation}
    k=\frac{1}{2}\overline{u'_i u'_i},
\end{equation}
and its transport can be written in the form used by Pope \cite{pope_turbulent_2000} as
\begin{equation}
    \underbrace{\overline{U_j}\frac{\partial k}{\partial x_j}}
       _{\substack{\displaystyle \mathcal{C} \\ \text{convection}}} 
    +\underbrace{\frac{1}{2}\frac{\partial}{\partial x_j} \overline{u'_i u'_i u'_j}}
       _{\substack{\displaystyle T^{(c)} \\ \text{turbulent} \\ \text{convection}}} \;
    +\underbrace{\frac{1}{\rho}\frac{\partial}{\partial x_j}\overline{u'_j p'}}
       _{\substack{\displaystyle T^{(p)} \\ \text{pressure} \\ \text{transport}}}
    -\underbrace{2\nu \frac{\partial}{\partial x_j}\overline{u'_i S'_{ij}}}
       _{\substack{\displaystyle T^{(\nu)} \\ \text{viscous} \\ \text{diffusion}}}
    = 
    \underbrace{- \overline{u'_i u'_j} \frac{\partial \overline{U_i}}{\partial x_j}}
       _{\substack{\displaystyle \mathcal{P} \\ \text{production}}} \;
    -\underbrace{2\nu \overline{S'_{ij}S'_{ij}}}
       _{\substack{\displaystyle \varepsilon \\ \text{viscous} \\ \text{dissipation}}}
    \label{eq:tke}
\end{equation}
\noindent where $p'$ is the fluctuating pressure, and $S'_{ij}$ is the fluctuating strain rate tensor $S'_{ij}=\frac{1}{2} \left( \frac{\partial u'_i}{\partial x_j} + \frac{\partial u'_j}{\partial x_i}\right)$. The three-dimensional nature of the flow requires that each term in Equation \ref{eq:tke} be summed over indices $j=1$, 2, and 3.

\subsection{Model Description}
The baseline whisker model is constructed using the geometric framework and dimensions presented by Hanke et al. \cite{hanke_harbor_2010}. There is variation in whisker dimensions within the harbor seal species {(\it{Phoca vitulina})} and at the individual level. Published average measurements also deviate slightly from one another as discussed in the review by Zheng et al. \cite{zheng_creating_2021}. While measurements have been reported by several sources including Hanke et al.\cite{hanke_harbor_2010}, Ginter et al. \cite{ginter_fused_2012}, Rinehart et al. \cite{rinehart_characterization_2017}, and Murphy et al. \cite{murphy_effect_2013}, the model and measurements presented by Hanke et al. comprised of average dimensions obtained through photogrammetry of 13 whiskers remains one of the standard representations of harbor seal whisker geometry \cite{witte_wake_2012,hans_whisker-like_2013,wang_wake_2016,yoon_effect_2020}. However, this model does not enable easy examination of geometric modifications as the defined dimensions are coupled with one another. Therefore, the model parameters are redefined in terms of hydrodynamic relevance and nondimensionalized to allow each parameter to be varied independently of one another. Descriptions of the model parameters are detailed by Lyons et al. \cite{lyons_flow_2020}.

\begin{figure}
    \centerline{\includegraphics[width=3.8in]{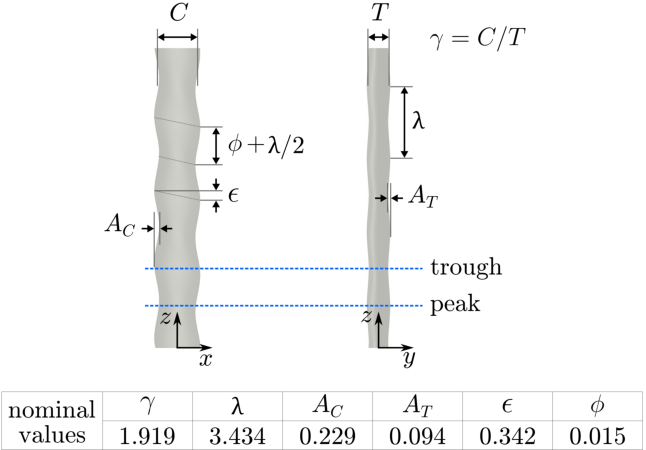}}
    \caption{Top and side view of the seal whisker model with nondimensional geometric parameters identified.}
    \label{fig:parameters}
\end{figure}

Figure \ref{fig:parameters} displays the seal whisker geometry with average chord length $C$, average thickness $T$, and undulation amplitudes in the chord and thickness, $A_C$ and $A_T$, respectively. The periodicity of the topography is governed by the wavelength ($\lambda$) and is further perturbed by the undulation asymmetry ($\phi$) and offset ($\epsilon$) parameters. The nominal values for the baseline harbor seal model presented by Hanke et al. \cite{hanke_harbor_2010} are listed at the bottom of figure \ref{fig:parameters}. The chord length and thickness are combined to form the aspect ratio ($\gamma$) and amplitudes $A_C$ and $A_T$ are nondimensionalized by the average thickness $T$, while $\phi$ and $\epsilon$ are nondimensionalized by wavelength. A detailed description of the conversion of geometric parameters from the definition proposed by Hanke et al. \cite{hanke_harbor_2010} is included in work by Lyons et al. \cite{lyons_flow_2020}. In figure \ref{fig:parameters}, the spanwise locations of maximum and minimum thickness amplitude are marked with dotted lines and referred to hereafter as \textit{peak} and \textit{trough} respectively. 

\begin{figure}
    \centerline{\includegraphics[width=4.3in]{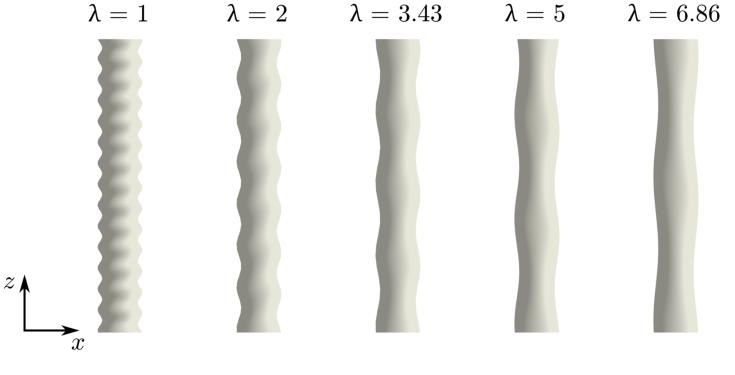}}
    \caption{Top view of whisker inspired models.}
    \label{fig:models}
\end{figure}

To investigate the effect of wavelength, five different wavelength seal whisker-inspired geometries are simulated and compared with a smooth elliptical cylinder of the same aspect ratio. The baseline model with $\lambda=3.43$ is simulated, along with two models with lower wavelength, $\lambda=1$ and $\lambda=2$, and two models with higher wavelength, $\lambda=5$ and $\lambda=6.86$. The top view of each model is shown in figure \ref{fig:models} with its associated wavelength value. The other five nondimensional parameters are held constant across the models.

\subsection{Computational Flow Parameters}
The drag and lift force coefficients for each geometry are calculated to gain an understanding of the bulk force response. Drag and lift coefficients are calculated from the drag force $F_D$ and lift force $F_L$ as
    \begin{align}
        \label{e:dragLift}
        C_D &= \frac {2F_D}{\rho {U^2_{\infty}} T L_z} \\[1.5ex]
            &\quad\text{and} \nonumber \\
        C_L &= \frac {2F_L}{\rho {U^2_{\infty}} C L_z} \, .
    \end{align}
The drag and lift coefficients are normalized by the average frontal area, $TL_z$, and average planform area, $CL_z$, respectively, where $L_z$ is the span. The lack of camber in all models produces a nominally zero value for $C_L$, thus the root-mean-square value, $C_{L,RMS}$, is used for analysis. The Strouhal number $St$ is defined as the dominant nondimensional frequency of the lift force spectrum $f^*$. The lift force frequency $f$ is nondimensionalized as
    \begin{equation}
        f^* = \frac{fT}{U_{\infty}} \, .
    \end{equation}

Calculation of the mean velocity $\overline{U_j}$ and mean pressure fields are computed over 600 nondimensional convective time units, $t^*=tU_{\infty}/T$. Spatial derivatives are calculated within \textit{OpenFOAM} on the original computational mesh. Fluctuating components are calculated for 301 unique time instances spanning $30t^*$ (equivalent to between 4.4 and 7.7 shedding cycles depending on the model). The use of $30t^*$ is determined to be sufficiently large to achieve convergence of time-averaged fluctuating terms. A negligible difference is seen between calculations using $20t^*$ and $30t^*$ for the $\lambda=3.43$ case. After calculation, all fields are sampled onto a coarser three-dimensional Cartesian grid (resolution of $\Delta x=\Delta y=\Delta z=0.1T$) to enable further post-processing and import into \textit{Matlab} where terms are time-averaged and visualized.

\section{Results and Discussion}\label{sec:results}
\subsection{Effect of Wavelength on Forces and Flow Structures}
\begin{figure}
    \centerline{\includegraphics[width=\textwidth]{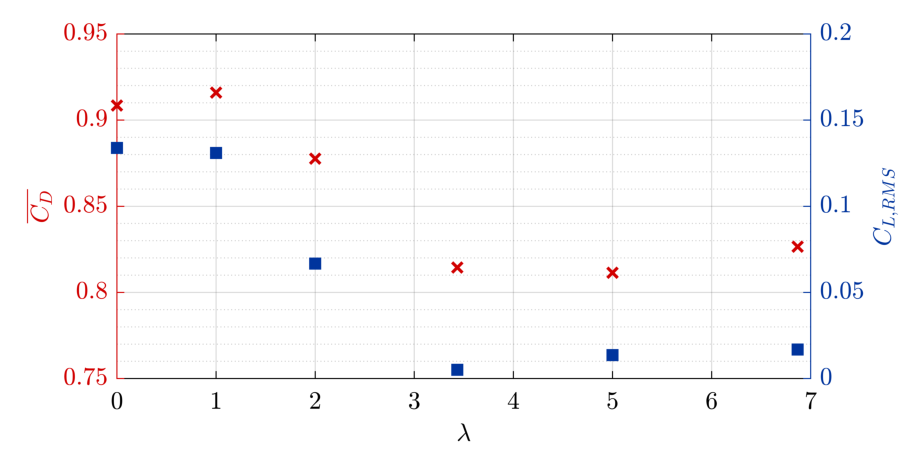}}
    \caption{For models with $\lambda>2$, there is a significant reduction in both $\overline{C_D}$ and $C_{L,RMS}$ when compared with the smooth ellipse case (plotted as $\lambda=0$). Both quantities $\overline{C_D}$ and $C_{L,RMS}$ sharply decrease to minima at $\lambda=5$ and $\lambda=3.43$ respectively before increasing again.}
    \label{fig:CdClplot}
\end{figure}

The plot of $\overline{C_D}$ and $C_{L,RMS}$ values in figure \ref{fig:CdClplot} illustrates the considerable range of oscillating lift and drag forces that result from wavelength variation. For ease of comparison, the ellipse is represented by a value of $\lambda=0$ on the $x$-axis. The baseline seal whisker geometry ($\lambda=3.43$) has 10\% lower drag and 96\% lower RMS lift than the ellipse. A similar reduction in bulk force response has been noted by others \cite{hanke_harbor_2010,witte_wake_2012,kim_effect_2017}. When $\lambda=1$, the force values are quite similar to those of the smooth ellipse even though other geometric modifications are present. However, the $\overline{C_D}$ value is slightly higher for this case than for the smooth ellipse, indicating that the introduction of undulations alone does not reduce drag, but undulation wavelength is important as well.

As wavelength increases to $\lambda=3.43$, $\overline{C_D}$ and $C_{L,RMS}$ decrease. At higher wavelengths, $\lambda=5$ and $\lambda=6.86$, $C_{L,RMS}$ increases slightly while $\overline{C_D}$ reaches its minimum at $\lambda=5$. Witte et al. \cite{witte_wake_2012} also simulated a whisker variation with a twice-nominal wavelength ($\lambda=6.86$), and their results show similar drag but lower oscillating lift forces. The discrepancy may be due to setting the undulation offset $\epsilon$ to zero in their models. The trend of lift and drag forces with respect to wavelength shown in figure \ref{fig:CdClplot} is similar to the non-linearity seen for wavy cylinder wavelength variations \cite{lam_effects_2009} and helically twisted elliptical cylinders \cite{kim_flow_2016}. 

\begin{figure}
    \centerline{\includegraphics[width=\textwidth]{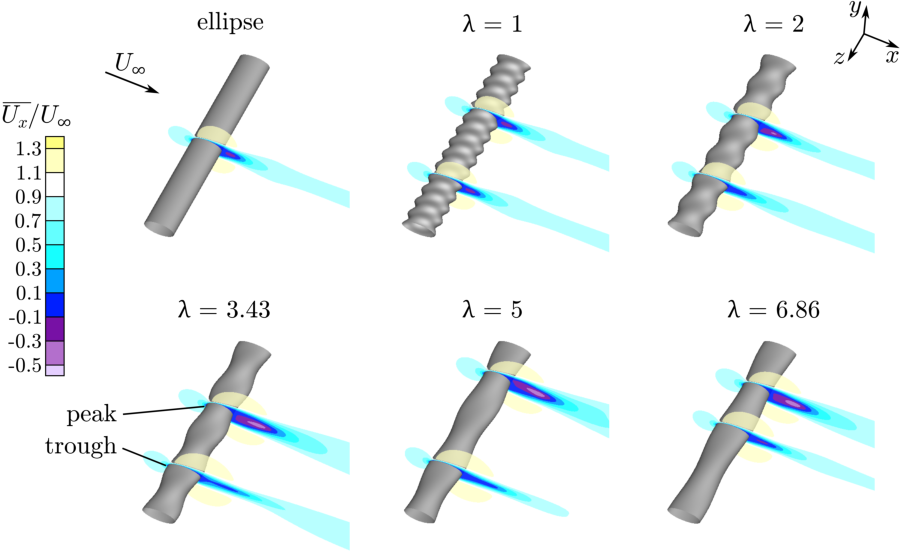}}
    \caption{Contours of mean streamwise velocity display the variation within the recirculation region as a function of wavelength. Contours are displayed at a representative peak and trough cross-section for each model.}
    \label{fig:UMeanContours}
\end{figure}

As a complement to the bulk force response, figure \ref{fig:UMeanContours} displays contours of time-averaged streamwise velocity $\overline{U_x}/U_{\infty}$ at peak and trough cross-sections for each model. Flow variations are solely due to changes in the frequency of undulation, as the geometric cross-sections at respective peaks and troughs are identical. The peak and trough velocity profiles for the $\lambda=1$ case are similar to those of the smooth ellipse as might be expected from the similarity in force response. Both cross-sections of the $\lambda=2$ case display a longer recirculation length than the ellipse and $\lambda=1$ models, and the peak cross-section has a larger area of reversed flow while flow behind the trough cross-section is nominally in the positive $x$-direction. The $\lambda=3.43$, 5, and 6.86 cases have a larger variation between their peak and trough cross-sections. For each, the trough cross-section has minimal flow reversal indicating a reduction in shear layer roll up.

\begin{figure}
    \centerline{\includegraphics[width=\textwidth]{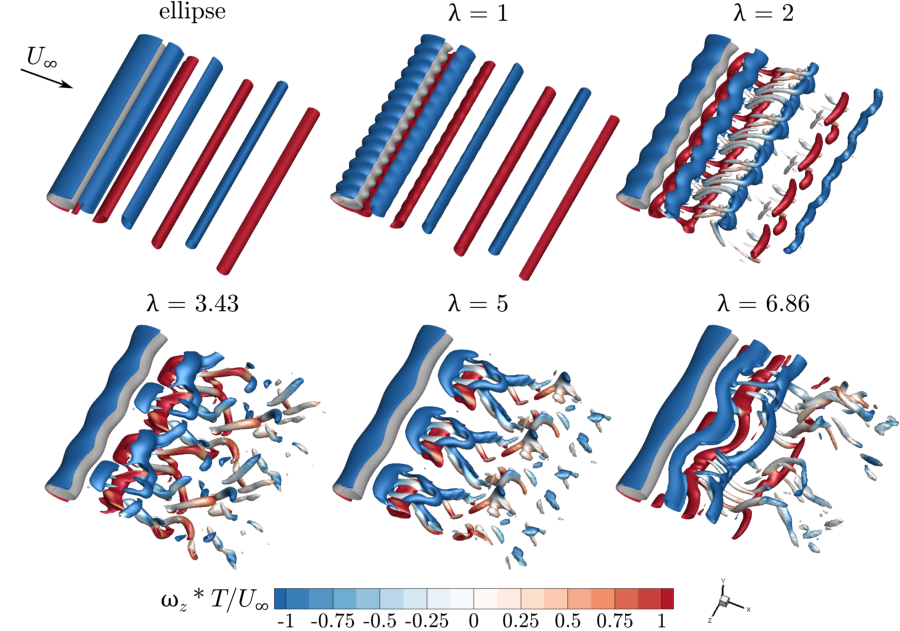}}
    \caption{Isosurfaces of nondimensional $Q=0.3$ colored by $z$-vorticity illustrate the effect of wavelength modification on flow structure patterns. Flow over the ellipse and $\lambda=1$ are primarily two-dimensional, while the $\lambda=2$ model generates secondary $x$- and $y$-vorticity, and hairpin vortices are visible at larger wavelengths.}
    \label{fig:Qisosurfaces}
\end{figure}

The variation in the mean velocity profiles indicates that changes in undulation frequency result in flow pattern variation. To visualize flow structures, isosurfaces of nondimensional $Q$ are plotted in figure \ref{fig:Qisosurfaces} and colored by $z$-vorticity with red indicating positive and blue indicating negative vorticity. The shed structures behind the $\lambda=1$ and 2 models are similar to a typical von K\'{a}rm\'{a}n street. However, the $\lambda=1$ model produces nominally two-dimensional vortices, whereas flow over the $\lambda=2$ model includes additional waviness in the spanwise vortex rolls, and braid-like structures form between rolls. These three-dimensional braid-like structures are created as streamwise and transverse vorticity develop. Secondary vorticity develops naturally in the flow behind smooth circular cylinders as a result of deformation of primary vortex cores leading to Mode-A instability \cite{leweke_three-dimensional_1998,williamson_vortex_1996,williamson_existence_1988,roshko_perspectives_1993}. However, the secondary structures observed for the $\lambda=2$ model are initiated by the increased spanwise velocity induced by the undulations of the model and are more closely spaced along the span than three to four diameters as typically seen for Mode-A \cite{williamson_existence_1988}.

In the wake of a smooth circular cylinder at similar $\Rey$, it is common for the flow to develop vortex dislocations along the spanwise length in addition to Mode-A instability \cite{gerrard_wakes_1978,williamson_natural_1992}. These intermittent occurrences lead to a break in the periodicity of Mode-A instability and typically cause a large decrease in the drag force time history \cite{behara_wake_2010}. Examination of force data indicates no such intermittent variations for any of the whisker-inspired models at this $\Rey$. Flow over the $\lambda=3.43$ and $\lambda=5$ models is periodic in time as shown by the lift force and three-dimensional with hairpin vortex structures developing in the wake. The $\lambda=5$ case develops vortex structures that are shed in tandem along the span. However, the structures behind $\lambda=3.43$ alternate shedding from top and bottom at 180 degrees of phase for each wavelength section along the span resulting in a near zero instantaneous $C_L$ as upper and lower forces work to negate one another. For both of these cases, the undulations are spaced far enough apart that the spanwise coherent vortex structures do not interfere with one another as they develop and convect downstream. 
The wavy cylinder geometry also displays three-dimensional flow structure distortion due to changes in wavelength \cite{lam_effects_2009,zhang_piv_2005}. At wavelengths where forces were minimal, Lam \& Lin \cite{lam_effects_2009} hypothesize that the introduction of additional streamwise vorticity tends to stabilize the two-dimensional spanwise vorticity of the free shear layers and thus prevent roll-up.
The $\lambda=6.86$ case contains a combination of patterns. Hairpin-like features are visible downstream of peak cross-sections, while largely two-dimensional roller-type structures appear downstream of the trough. The larger distance between undulations, and subsequently between hairpin structures, enables the development of additional vortex rollers. In contrast, the close spanwise spacing of the hairpin vortices in the $\lambda=3.43$ and 5 cases prevents shear layer roll-up behind trough regions of the geometry.

\subsection{Analysis of Turbulence Kinetic Energy}
\begin{figure}
    \centerline{\includegraphics[width=\textwidth]{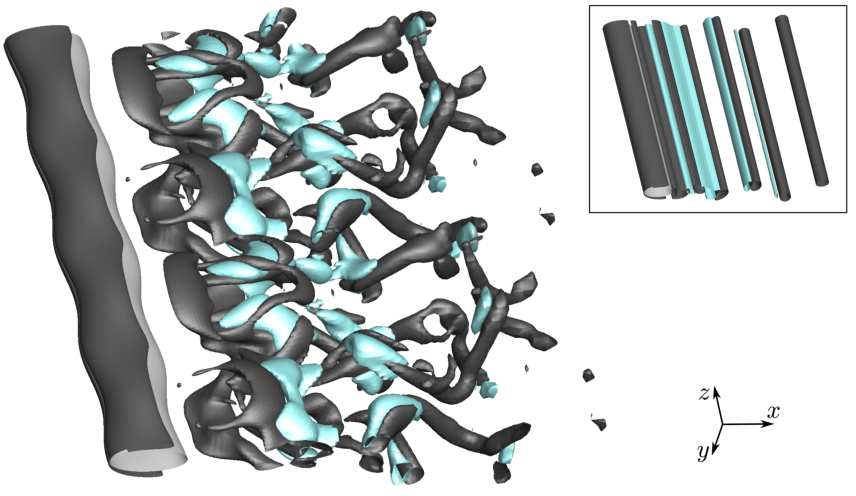}}
    \caption{Isosurfaces of nondimensional $Q=0.3$ shown in gray and instantaneous turbulence kinetic energy $k'=0.08$ in blue for the $\lambda=3.43$ case highlight the link between vortex flow structures and TKE. For comparison, the ellipse case is displayed in the upper right with isosurface values of nondimensional $Q=0.3$ in gray and $k'=0.12$ in blue.}
    \label{fig:Q_tke_isosurfaces}
\end{figure}

The instantaneous flow structures illustrated by isosurfaces of $Q$-criterion offer insight into the development and transfer of TKE. Isosurfaces of $Q$ for $\lambda=3.43$ are repeated in figure \ref{fig:Q_tke_isosurfaces} (gray) and superimposed with isosurfaces of instantaneous turbulence kinetic energy,
\begin{equation}
    k'=\frac{1}{2}\,u'_iu'_i,
\end{equation}
\noindent shown in blue. A moderate $k'$ isosurface level is chosen in order to depict structures which envelope regions of higher $k'$. The $k'$ isosurfaces are interlaced with those of $Q$ and convect downstream with the vortex structure. Regions of large $k'$ form as shear is generated between vortices of different rotation, enabling energy transfer from the mean flow to turbulence. In the upper right of figure \ref{fig:Q_tke_isosurfaces}, isosurfaces of nondimensional $Q$ and $k'$ in the wake of a smooth ellipse are displayed for comparison.

\begin{figure}
    \centerline{\includegraphics[width=\textwidth]{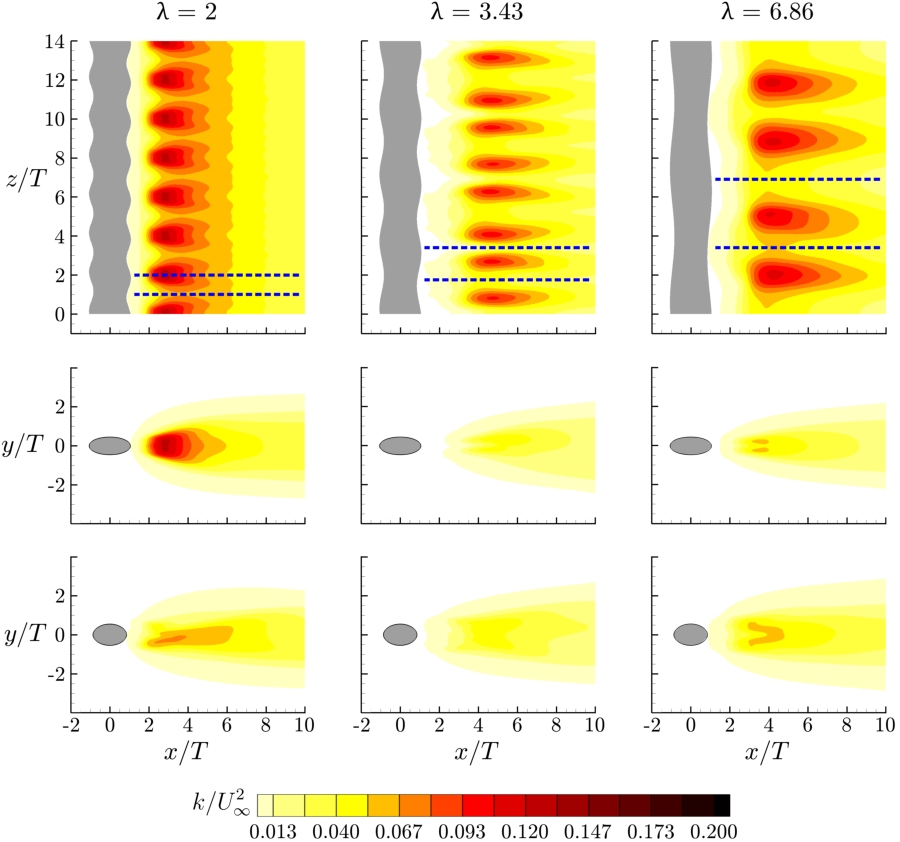}}
    \caption{Contours of TKE for $\lambda=2$, 3.43, and 6.86. In the top row, slices at $y=0$ display the spanwise periodic pattern of TKE and dashed lines are shown at trough and peak locations. Slices in the $x$-$y$ plane at trough and peak locations, in rows two and three respectively, highlight the difference in TKE due to undulation wavelength.}
    \label{fig:tkeContours}
\end{figure}

As energy is transferred from the mean flow into TKE, inspection of TKE contours allows for a more detailed comparison between models. The top row in figure \ref{fig:tkeContours} displays contours of TKE in the $x$-$z$ plane at $y=0$. The cases $\lambda=2$, 3.43, and 6.86 are chosen to illustrate the effects representative of a small, medium, and large wavelength as the $\lambda=1$ case has been shown to exhibit characteristics most similar to a smooth elliptical cylinder. In each case, the largest values of TKE appear periodically along the span with a frequency dependent on the undulation wavelength. Maximum TKE values occur downstream of the recirculation region, thus the location of maximum TKE is further downstream for the $\lambda=3.43$ and $\lambda=6.86$ cases than for $\lambda=2$, mirroring the pattern seen in the elongation of the recirculation lengths in figure \ref{fig:UMeanContours}. Furthermore, the maximum magnitude of TKE is noticeably lower for these larger wavelength models. The lower TKE values in the near wake correlate with lower $\overline{C_D}$ and $C_{L,RMS}$ as indicated in figure \ref{fig:CdClplot}. This is consistent with the direct relationship between TKE and lift and drag forces previously noted by Yoon et al. \cite{yoon_effect_2020} and Chu et al. \cite{chu_three-dimensional_2021} for whisker-inspired geometries and by Lin et al. \cite{lin_effects_2016} and Zhang et al. \cite{zhang_piv_2005} for wavy cylinders.

The bottom two rows in figure \ref{fig:tkeContours} contain $x$-$y$ cross-sections at trough and peak locations. The cross-sectional views show the maximum TKE values for $\lambda=3.43$ and $\lambda=6.86$ occur neither at the trough nor peak cross-section, but rather in between, a pattern also noted by Yoon et al. \cite{yoon_effect_2020}. Conversely, the $\lambda=2$ case contains a maximum behind the trough cross-section, similar to wavy cylinders of similar wavelength where TKE maxima appear and streamwise velocity recovers more quickly behind node locations \cite{zhang_piv_2005}. The introduction of hairpin vortex structures at wavelengths larger than 2 may initiate the spanwise shift of the TKE maxima as the shear layer is more stable at trough locations for these geometries.

\begin{figure}
    \centerline{\includegraphics[width=\textwidth]{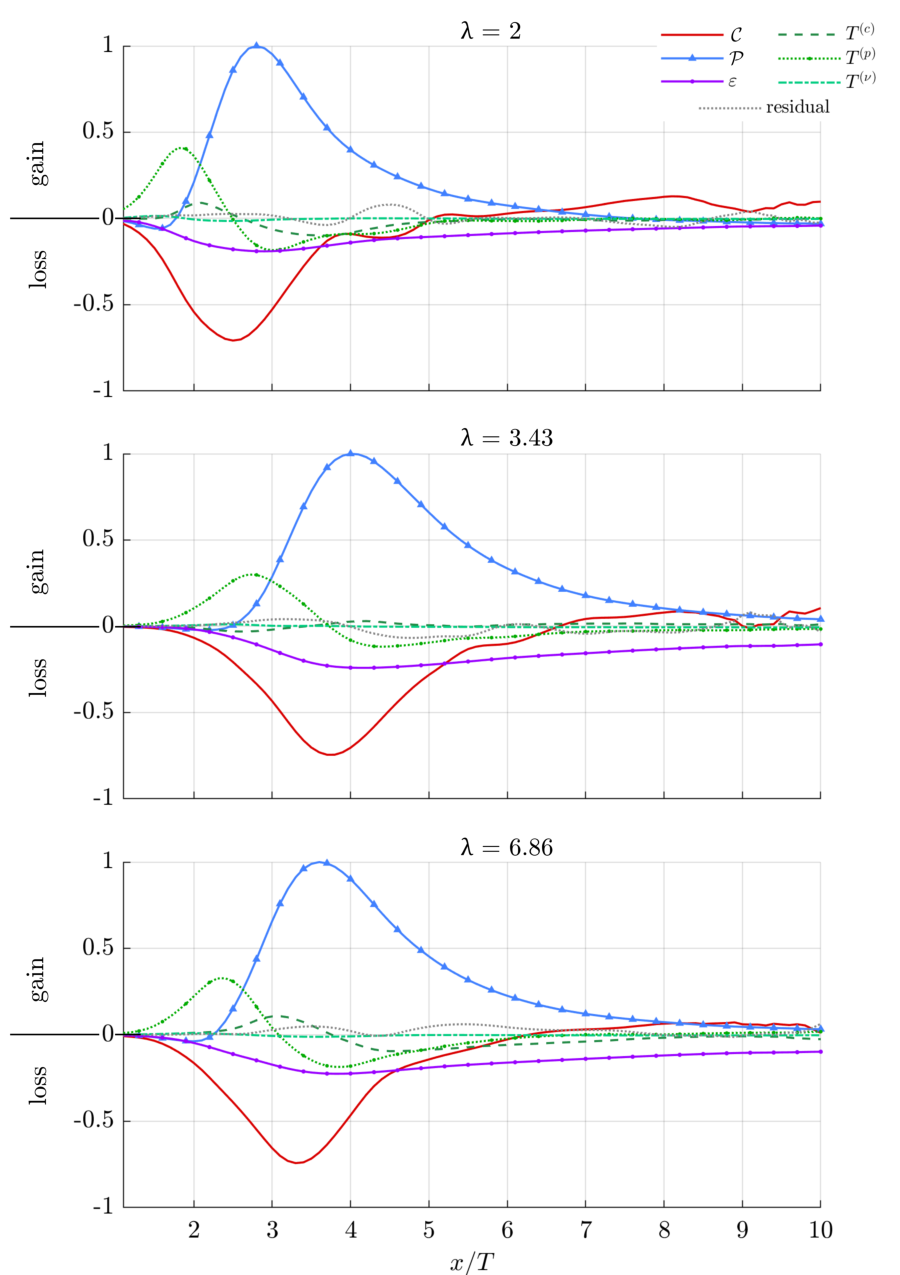}}
    \caption{Budgets of TKE transport with respect to downstream location. Values are averaged in the $y$-direction from $y/T=0$ to 2 as well as span-averaged.}
    \label{fig:yAvgBudgets}
\end{figure}

While force trends and general flow structures largely agree with those previously investigated for whisker-inspired geometries \cite{hanke_harbor_2010,witte_wake_2012,kim_effect_2017}, wavy cylinders \cite{lam_effects_2009,zhang_piv_2005}, and helically twisted elliptical cylinders \cite{kim_flow_2016}, calculation and analysis of TKE transport terms for flow over a seal whisker geometry are completed in the next sections for the first time to the authors' knowledge. The relationship among the TKE transport terms with respect to downstream development is illustrated in figure \ref{fig:yAvgBudgets}. To gain an understanding of the overall transport trend with respect to the streamwise direction alone, the TKE equation terms are averaged over a full wavelength span from $\lambda/2$ to $3\lambda/2$ and averaged over a domain of $y/T=$ 0 to 2. The three plots are normalized by the maximum average production for each associated case to allow for easier comparison. The turbulent transport terms, $T^{(c)}$, $T^{(p)}$, and $T^{(\nu)}$, are represented by various dashed green lines, while mean convection, $\mathcal{C}$, is designated by a solid red line. The peak average production shifts further downstream for the larger wavelength models. Likewise, peak values for $\mathcal{C}$ and $T^{(p)}$ occur further downstream for $\lambda=3.43$ and 6.86. 

Peak $T^{(p)}$ values appear within the average recirculation region for each model, a typical location for similar geometries such as a circular cylinder \cite{tian_new_2020}. Immediately downstream of the body, $\mathcal{P}$ is negative indicating that the average flow in this region is dominated by reversed flow and TKE is transferred back to the mean flow. For the $\lambda=2$ case, peak $T^{(p)}$ occurs at the downstream location where $\mathcal{P}$ first crosses zero. In comparison, the $T^{(p)}$ curve reaches its peak further downstream after $\mathcal{P}$ is already positive for the $\lambda=3.43$ and 6.86 geometries. At its peak, $\mathcal{P}$ is balanced predominately by mean convection, dissipation, and pressure transport. While the magnitude of mean convection is similar for all three cases, the proportion of $T^{(p)}$ and $\varepsilon$ varies. The ratio $T^{(p)}/\varepsilon$ for $\lambda=2$ and 6.86 is 0.82 and 0.75, respectively, whereas the ratio for $\lambda=3.43$ is much lower at 0.34. The lower ratio for the 3.43 case indicates the tendency for the flow to dissipate turbulence in this region rather than redistribute it via fluctuating pressure. A decrease in pressure fluctuations, especially in the near wake, has implications for reduction of flow noise.

A detailed understanding of the effects of wavelength variation on the production and dissipation terms can be gained by examining them individually. The production term, $\mathcal{P}=-\overline{u'_iu'_j}\left({\partial \overline{U_i}}/{\partial x_j}\right)$, in the TKE transport equation represents the energy converted from mean kinetic energy into TKE through mean shear interaction with Reynolds stresses. In this way, TKE provides a connection between the mean and turbulent quantities. The location and magnitude of $\mathcal{P}$ are dependent on the mean shear and turbulent fluctuations induced by near wake flow structures. 

\begin{figure}
    \centerline{\includegraphics[width=\textwidth]{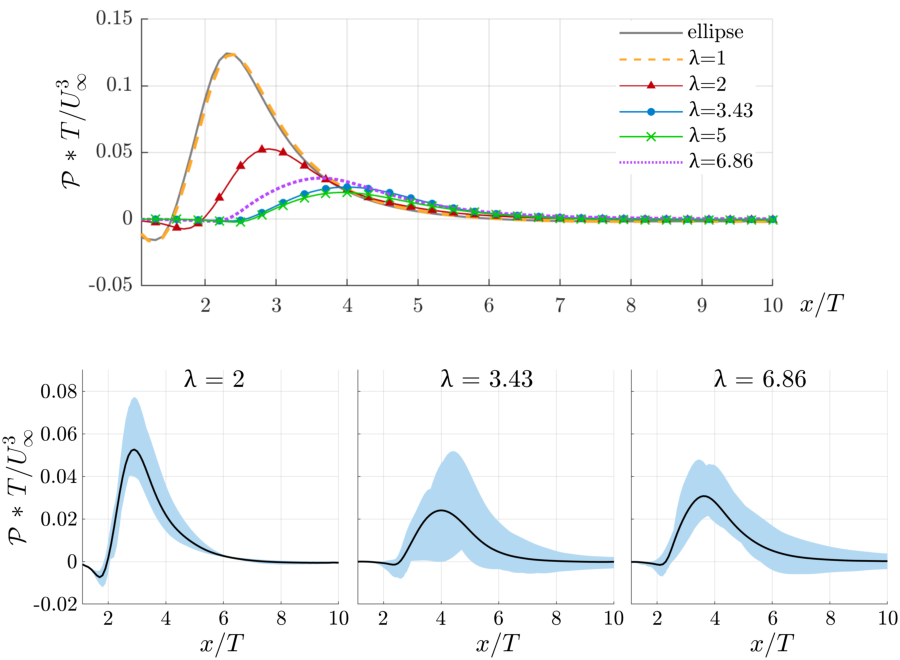}}
    \caption{Span-averaged curves of production along the wake centerline $y/T=0$. Below, spanwise variation of wake centerline production values are shaded in blue with the span-averaged quantity shown as a solid line.}
    \label{fig:prodComparison}
\end{figure}

In order to compare the models of varying wavelength, the production terms at the wake centerline $y/T=0$ are span-averaged and are displayed in figure \ref{fig:prodComparison}. Similar to the force response, the $\lambda=1$ model closely follows the ellipse production trend line. For the $\lambda=2$ geometry, the peak production occurs further downstream and with a lower maximum value. The flow over $\lambda=3.43$ and 5 have yet lower peak production values with occurrence further downstream. Finally, the $\lambda=6.86$ flow reverses the decreasing trend with slightly larger peak production occurring earlier than $\lambda=3.43$.

The spanwise variation of the wake centerline production is illustrated in the bottom portion of figure \ref{fig:prodComparison} by the blue shading for $\lambda=2$, 3.43, and 6.86 while the solid line represents the same span-averaged values as in the upper portion. The variation with respect to spanwise position provides a more detailed picture of the production than the span-averaged curves that reduce the flow to one dimension. The peak production for $\lambda=2$ occurs earlier than the other two cases and with considerably less variation. Centerline production for $\lambda=3.43$ has the largest variation with local zero values possible even as the span-average reaches its peak. Both $\lambda=3.43$ and 6.86 have considerable variability $3T$ downstream of peak production while downstream production from the $\lambda=2$ case remains nominally zero.

\begin{figure}
    \centerline{\includegraphics[width=\textwidth]{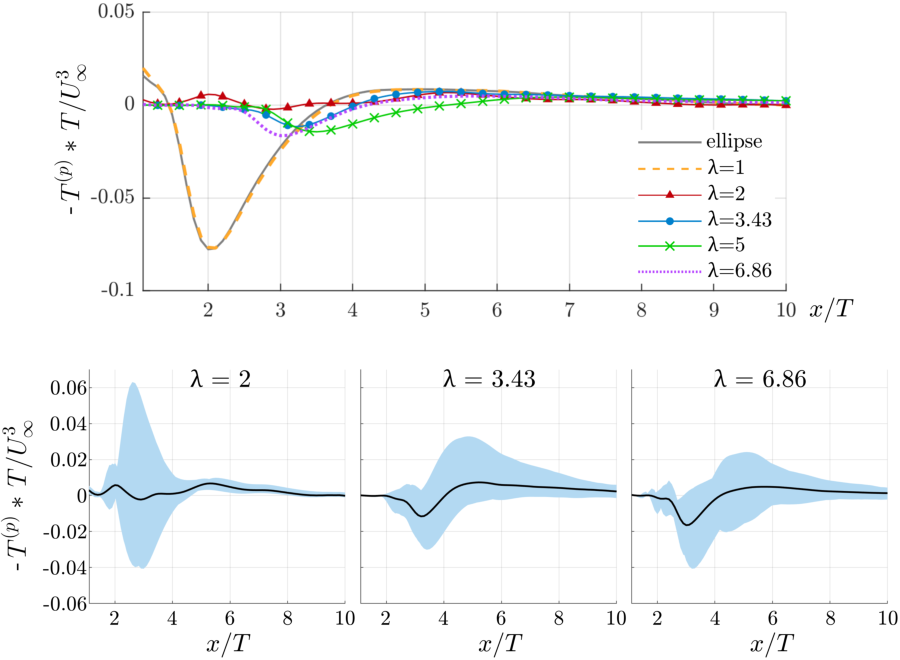}}
    \caption{Span-averaged curves of pressure transport along the wake centerline $y/T=0$. Below, spanwise variation of wake centerline pressure transport values are shaded in blue with the span-averaged quantity shown as a solid line.}
    \label{fig:TpComparison}
\end{figure}

As shown in figure \ref{fig:yAvgBudgets}, one of the primary transport mechanisms in the near wake region is the turbulent pressure transport $T^{(p)}$, which redistributes energy among the TKE equation terms. The span-averaged pressure transport values are plotted with respect to downstream location at $y/T=0$ in figure \ref{fig:TpComparison}. For $\lambda>2$, the overall magnitude of the span-averaged pressure transport is considerably reduced compared to the large peak present in the elliptical cylinder case. In the lower portion of the figure, the spanwise variation is shaded in blue. While span-averaged values are minimal, the variation along the span is substantial, indicating the importance of local three-dimensional effects. Larger wavelengths ($\lambda=3.43$ and 6.86) extend the range of these spanwise effects further downstream as indicated by the larger variation for $x/T>4$ compared to the small variation seen for the $\lambda=2$ case.

\begin{figure}
    \centerline{\includegraphics[width=\textwidth]{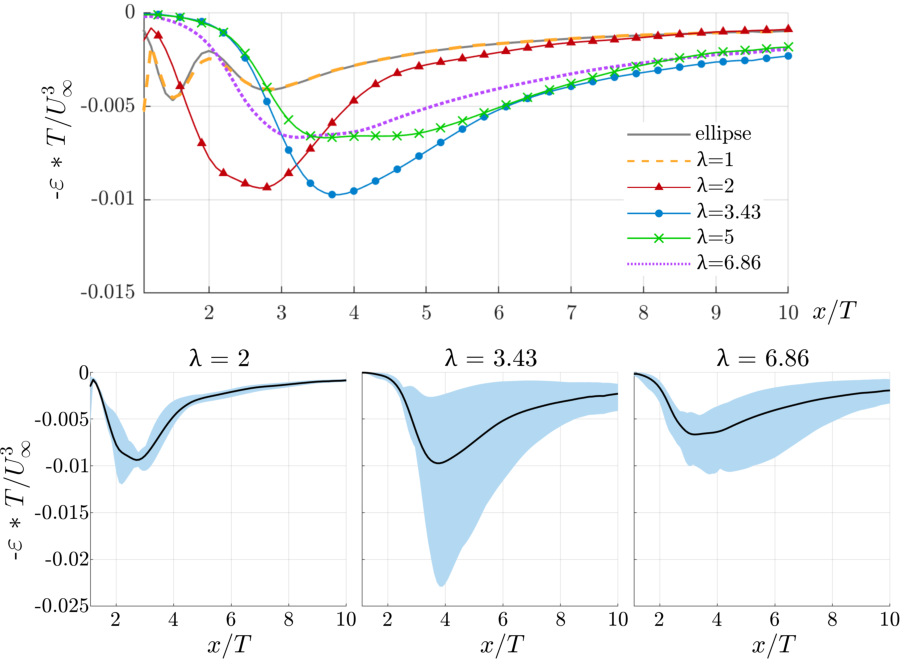}}
    \caption{Span-averaged curves of dissipation along the wake centerline $y/T=0$. Below, spanwise variation of wake centerline dissipation values are shaded in blue with the span-averaged quantity shown as a solid line.}
    \label{fig:epsilonComparison}
\end{figure}

In primary opposition to $\mathcal{P}$, removal of TKE is accomplished predominantly through viscous dissipation, $\varepsilon = 2\nu \overline{S'_{ij}S'_{ij}}$, at the smallest length scales. Governed by the fluctuating strain rate, dissipation is largest where fluctuating terms are located. Not all cases achieve peak dissipation along the wake centerline $y/T=0$, nevertheless, the span-averaged wake centerline dissipation values provide ease of comparison and are plotted in figure \ref{fig:epsilonComparison}. Both $\lambda=2$ and 3.43 have large dissipation peaks although the $\lambda=2$ peak occurs closer to the body. In contrast, $\lambda=5$ and 6.86 have broad, flat peaks between $x/T=3$ and 5 rather than a sharp peak as seen for $\lambda=2$ and 3.43. 

Spanwise variation of centerline dissipation is shown by the blue shading in the lower portion of figure \ref{fig:epsilonComparison} where the solid black line is the span-averaged value. Although $\lambda=2$ and 3.43 have peak span-averaged centerline dissipation values of similar magnitude, there is little spanwise variation in $\lambda=2$ whereas the $\lambda=3.43$ flow has a wide range of dissipation values and considerable dissipation at further downstream locations ($x/T >6$). The wake of $\lambda=6.86$ maintains dissipation variation downstream as well, but to a lesser extent. Both the lower level of TKE production and the higher amount of dissipation contribute to the $\lambda=3.43$ flow having the lowest levels of TKE compared to the other wavelengths.

\subsection{Turbulence Kinetic Energy Spanwise Effects}
\begin{figure}
    \centering
    \subfloat[Contours of production.]{
        \centerline{\includegraphics[width=\textwidth]{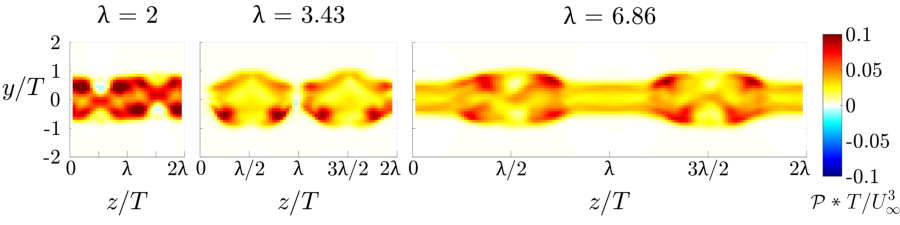}}
        \label{fig:prodContours}
        }
        
    \subfloat[Contours of dissipation.]{
        \centerline{\includegraphics[width=\textwidth]{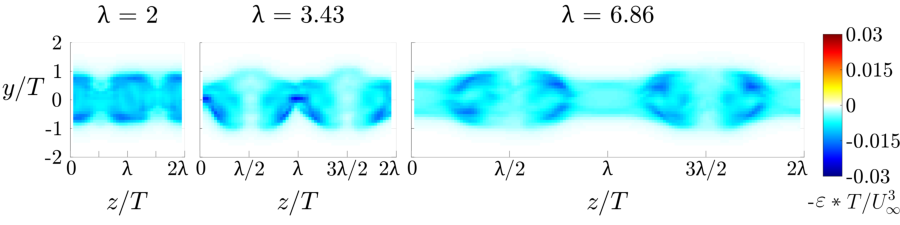}}
        \label{fig:epsilonContours}
        }
    \caption{Contours of production and dissipation in the $y$-$z$ plane are displayed at locations of maximum span-averaged production (as displayed in figure \ref{fig:prodComparison}), $x/T=2.9$, 4, and 3.6 for the $\lambda=2$, 3.43, and 6.86 models respectively.}
    \label{fig:contourSlices}
\end{figure}

Insight into the relationship between flow structure and $\mathcal{P}$ is gained by extracting a $y$-$z$ contour slice at the location of peak span-averaged production along the wake centerline $y/T=0$, profiles of which are shown in figure \ref{fig:prodComparison}. Figure \ref{fig:prodContours} displays contours of $\mathcal{P}$ at $x/T=2.9$, 4, and 3.6 for the $\lambda=2$, 3.43, and 6.86 flows, respectively. The shapes of the contours mimic the shapes of the vortical structures shown in figure \ref{fig:Qisosurfaces} with the largest values of $\mathcal{P}$ occurring near the edges of the vortex structures where mean shear interacts with Reynolds stresses. The wavy nature of the vortex rolls produced in the wake of $\lambda=2$ is visible in the periodic hot-spot pattern of the $\mathcal{P}$ contours, and the coherent hairpin structures that develop behind the $\lambda=3.43$ case create two distinguishable $\mathcal{P}$ patches centered at $z/T=\lambda/2$ and $3\lambda/2$. In this case, the two patches are separated with a small region of near-zero production occurring around $z/T=\lambda$. The combination of flow structures that develop behind the $\lambda=6.86$ geometry are apparent in the $\mathcal{P}$ contour pattern as well. Patches of production due to hairpin-like structures form near $z/T=\lambda/2$ and $3\lambda/2$, similar to the $\lambda=3.43$ case. However, a stripe pattern on either side of these patches appears resulting from the additional vortex rollers that appear at this wavelength. The combination of flow structures ultimately results in a higher overall level of $\mathcal{P}$ for the larger wavelength case than for $\lambda=3.43$ where secondary roller structures are suppressed. Likewise, large values of turbulent pressure transport are a result of the product of velocity and pressure fluctuations. Locations of large $T^{(p)}$ magnitude align generally with areas of high production and are not displayed here.

Contour slices of $-\varepsilon$ in the $y$-$z$ plane are taken at the same downstream locations and displayed in figure \ref{fig:epsilonContours}. The overall dominant shapes are similar to those of $\mathcal{P}$ as they both are linked to the underlying vortex structure, however, there are several notable differences. While both production and dissipation occur at locations where turbulent fluctuations are large, production of TKE requires mean strain while dissipation relies on fluctuating strain rates. Larger $\mathcal{P}$ values are located near shear layers, especially apparent by the horizontal stripe pattern shown in figure \ref{fig:prodContours} for the $\lambda=6.86$ case, whereas dissipation is more dispersed around the edges of vortex structures. Furthermore, the $\lambda=3.43$ case develops a large dissipative sink between the two shed structures at $z/T=\lambda$, and although space exists between the shed hairpin-like structures of the $\lambda=6.86$ case, a similar TKE sink is not found.

Viewing the production and dissipation contours in the $y$-$z$ plane also displays the asymmetry with respect to $y$ that occurs for the $\lambda=3.43$ flow. Larger $\mathcal{P}$ and $-\varepsilon$ values are located at negative $y$, while placement is more uniform for other wavelength models. It is possible that shed vortex structure with the inclusion of the hairpin vortex allows for increased stability in this orientation and this low $\Rey$.

\begin{figure}
    \centerline{\includegraphics[width=\textwidth]{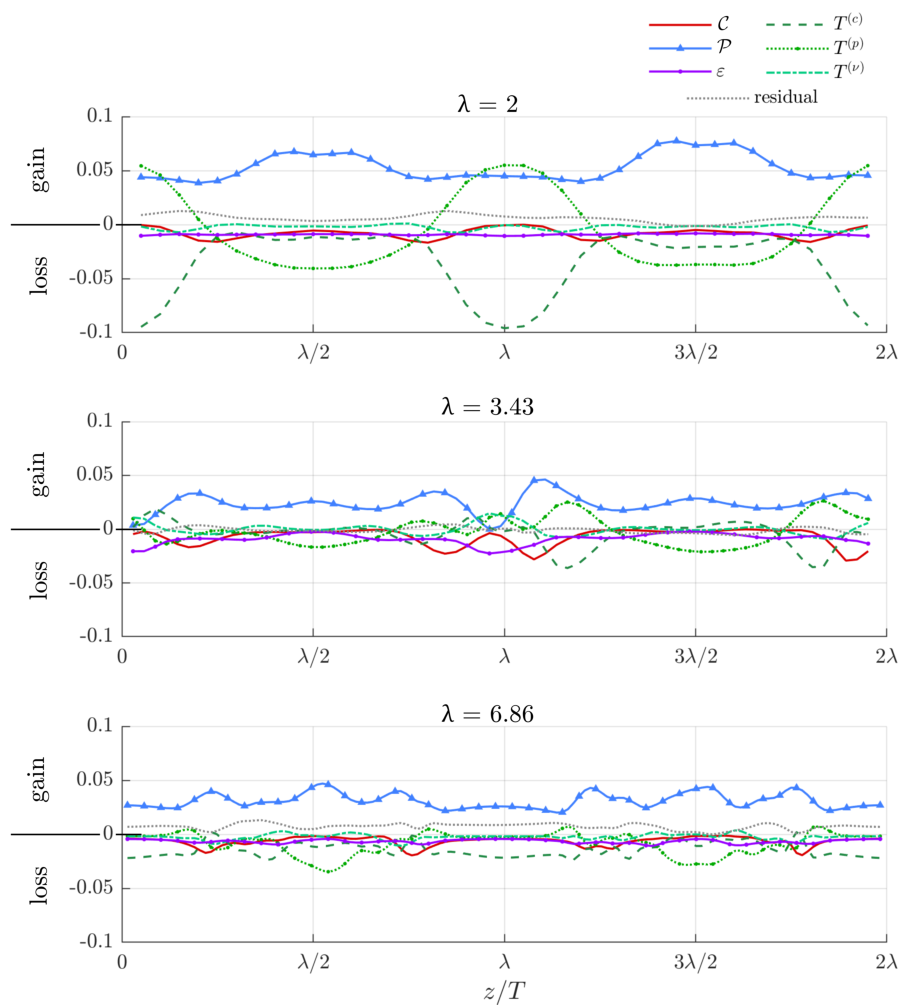}}
    \caption{Spanwise budgets of TKE transport along the wake centerline $y/T=0$ for $\lambda=2$, 3.43, and 6.86 cases at peak production downstream locations $x/T=2.9$, 4, and 3.6, respectively.}
    \label{fig:spanwiseBudgets}
\end{figure}

To examine the complexity of spanwise effects on TKE, the TKE budget along the wake centerline $y/T=0$ is shown as a function of spanwise position in figure \ref{fig:spanwiseBudgets} for $\lambda=2$, 3.43, and 6.86 at a downstream location of peak span-averaged production. Corresponding to the structures seen in figure \ref{fig:prodContours}, all three cases have local minima near $z/T=\lambda$. While the production term for the $\lambda=2$ and 6.86 models decreases moderately at $z/T=\lambda$, it diminishes to near zero around $\lambda$ for the $\lambda=3.43$ case as the proximity of the coherent hairpin structures prevents the development of oscillating vortex shedding in between. While dissipation remains relatively constant over the span for $\lambda=2$ and 6.86, it displays a maximum magnitude at $z/T=\lambda$ for the $\lambda=3.43$ case. Complementing production and dissipation, the remaining TKE terms are responsible for transport, of which the turbulent transport terms contribute more than mean convection. For $\lambda=2$, the turbulent pressure transport $T^{(p)}$ is largely balanced by turbulent convection $T^{(c)}$ near $z/T=\lambda$. The appearance of $T^{(c)}$ correlates with the higher $C_{L,RMS}$ value that manifests for the $\lambda=2$ case. For the three cases shown, the viscous diffusion term $T^{(\nu)}$ is generally small, but interestingly becomes a dominant term near $z/T=\lambda$ in the $\lambda=3.43$ case.

\section{Conclusions}\label{sec:conclusion}
Flow over five seal whisker inspired geometries with various undulation wavelengths is simulated and compared with a smooth elliptical cylinder at $\Rey=250$. An analysis of TKE transport terms is performed to explore the underlying mechanisms linking flow structures, TKE values, and force reduction. The drag and oscillating lift response is significantly reduced for the undulated geometries with the $\lambda=3.43$ case achieving a reduction of 96\% in $C_{L,RMS}$ and 10\% in $\overline{C_D}$, agreeing with trends in prior research.

Furthermore, modification of undulation wavelength gives rise to a variety of flow patterns with hairpin vortex structures forming for models with $\lambda>2$, replacing the two-dimensional von K\'{a}rm\'{a}n street of smaller wavelengths. The resulting three-dimensional flow structures influence the creation of mean shear and $\Rey$ stresses which, in turn, impact the magnitude and location of TKE production and dissipation. Topographies that minimize alternating vortex shedding create an elongated recirculation region and move the location of maximum TKE production further downstream. Analysis of TKE transport terms shows TKE in the near wake region is dominated by production and dissipation while turbulent pressure redistributes the energy among the terms. 

Although these trends are seen at multiple wavelengths, they are most amplified for the $\lambda=3.43$ model, which generates hairpin vortices with spanwise spacing such that they do not interfere with one another yet are close enough to prevent additional shear layer roll up in between. Such structures decrease TKE production and create space for a TKE sink between them, decreasing overall TKE in the near wake and corresponding forces on the body. At the downstream location of maximum production, this model generates a wake where turbulent fluctuations are more likely to dissipate rather than redistribute TKE as shown by the low value of the ratio $T^{(p)}/\varepsilon$. For the $\lambda=3.43$ case, $T^{(p)}/\varepsilon=0.34$ while $\lambda=2$ and 6.86 have ratios closer to unity at 0.82 and 0.75 respectively. In summary, the vortex structures both lower the TKE production and increase the dissipation allowing for a lower overall level of TKE compared to the other models. 

Understanding the impact of wavelength modification on vortex structure and the subsequent effects on specific TKE terms provides a guide for the selection of bio-inspired geometry parameters. While this work concentrates on variations in wavelength, further investigation would explore the effects of other geometric parameters and extend to higher $\Rey$.

\backmatter

\bmhead{Acknowledgements}
{The authors kindly thank Dr. Christin Murphy for her guidance, knowledge, and insight into biological systems. This research was conducted using computational resources provided by the Center for Computation and Visualization at Brown University and the US Department of Defense HPC Modernization Program.}

\section*{Declarations}
\bmhead{Ethical Approval}
not applicable

\bmhead{Competing interests}
{The authors report no conflict of interest.}

\bmhead{Authors' contributions}
{K.L. and J.A.F. were responsible for conceptualization. All authors assisted with the development of methodology. K.L. completed analysis and created the initial draft manuscript. All authors assisted with revisions and editing.}

\bmhead{Funding}
{This work was supported by the National Science Foundation (J.A.F. and K.L., grant number CBET-2035789), (R.B.C., grant number CBET-2037582) under program manger Ron Joslin.}

\bmhead{Availability of data and materials}
Data sets are available upon request.

\begin{appendices}
\section{Mesh Resolution Study}\label{appA}
Generating a high-quality, structured mesh surrounding the whisker models is time-intensive and complex given the three-dimensional undulations. An improved method is introduced by Yuasa et al. \cite{yuasa_simulations_2022} that utilizes a smooth mesh morphing algorithm coupled to the flow solver. Geometric parameters are specified and obtained through an analytical expression of the whisker surface topography to actively morph from one geometric realization to the next. In addition to the time savings in meshing, the method presented by Yuasa et al. ensures repeatable realizations of the geometry by defining the surface topography with a complete analytical expression. For the following simulations, an initial three-dimensional structured mesh is generated for a smooth ellipse of the same aspect ratio and span as the desired model. The flow over the smooth elliptical cylinder is computed for 100 nondimensional convective time units, $t^*=tU_{\infty}/T$, while the flow structures in the wake are developed. Then over the next five $t^*$, the mesh is morphed into the desired topography. The flow is given another 95$t^*$ to fully transition before flow field statics are collected for the next 600$t^*$.

\begin{figure}
    \centerline{\includegraphics[width=5.5in]{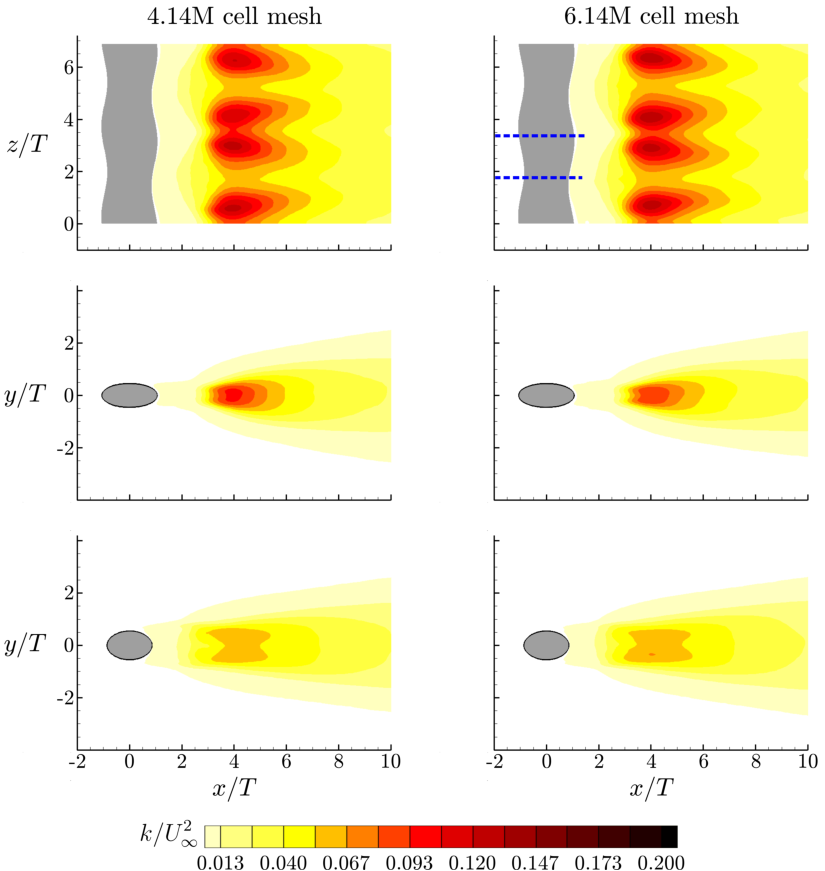}}
    \caption{Contours of TKE for the chosen mesh with 4.14M cells and a larger mesh with 6.14M cells show comparable results. The top row displays slices at $y=0$. Below, contours at trough and peak cross-sections are shown.}
    \label{fig:meshValidationContours}
\end{figure}

A comparison of mesh resolution results at $\Rey=500$ for the baseline $\lambda=3.43$ model is shown in table \ref{table:meshValidation}. While analysis is completed at $\Rey=250$, the mesh resolution is validated at the higher Reynolds number $\Rey=500$ to compare with data from prior literature. The table shows the value for $\overline{C_D}$ decreases slightly with increasing mesh size and converges to three decimal places for the two largest meshes. Similarly, $St$ remains constant for the three largest meshes. To compare performance in the wake, contours of TKE are plotted for the 4.14M cell and 6.14M cell meshes in figure \ref{fig:meshValidationContours}. Slices at $y=0$ are displayed in the top row showing comparable contour patterns and magnitudes. The dotted lines designate the locations of the trough and peak cross-sections shown in the second and third rows respectively. Slight differences can be seen in the $x$-$y$ cross-section contours at downstream locations ($x \gtrsim 4T$). The 4.14M cell mesh is chosen for analysis as further increases in mesh size show negligible differences.

\begin{figure}
    \subfloat[Resolution in near wake region]{
        \centerline{\includegraphics[width=4.5in]{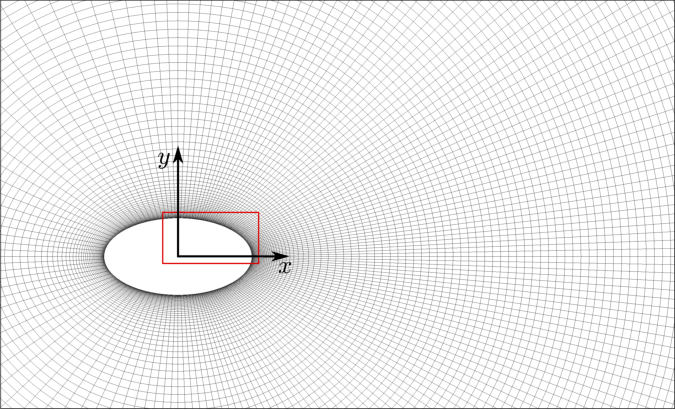}}
        \label{fig:mesh_a}
        }
        
    \subfloat[Zoomed view of figure \ref{fig:mesh_a} inset]{
        \centerline{\includegraphics[width=4.5in]{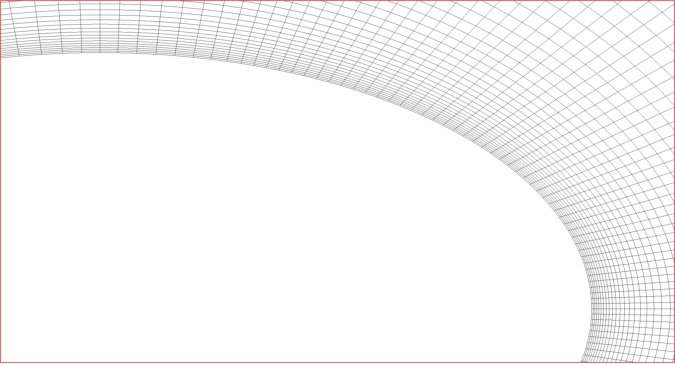}}
        \label{fig:mesh_b}
        }
    \caption{Mesh for $\lambda=3.43$ case contains 4.14M cells and provides adequate resolution near the wall and in the wake.}
    \label{fig:mesh}
\end{figure}

The mesh for the $\lambda=3.43$ is shown in figure \ref{fig:mesh_a} with an inset view of the near wall region in figure \ref{fig:mesh_b}. As wavelength is changed for each model, the whisker length is modified to maintain a two-wavelength spanwise domain. The number of mesh cells in the spanwise direction is adjusted accordingly for each model while the azimuthal and radial resolution is maintained.

{\rowcolors{2}{white}{lightgray!40}
\begin{table}
    \begin{center}
    \caption{Table includes the tested mesh resolutions for $\lambda=3.43$ case at $\Rey=500$ with the number of nodes in the spanwise ($N_z$), azimuthal ($N_{\theta}$), and radial ($N_r$) directions, the uniform spanwise spacing ($\Delta z/T$), and the minimum radial spacing ($\Delta r/T_{min}$). The 4.14M cell mesh is selected for use in further analysis.}
     \begin{tabular}{c c c c c c c c}
            \hline \hline
            $\boldsymbol{N_{total}}$ &
            $\boldsymbol{N_z}$ &
            $\boldsymbol{N_{\theta}}$ &
            $\boldsymbol{N_r}$ &
            $\boldsymbol{\Delta z/T}$ &
            $\boldsymbol{\Delta r/T_{min}}$ &
            $\boldsymbol{\overline{C_D}}$ &
            $\boldsymbol{St}$ \\
            \hline \hline
            1.01M  & 110 & 100 & 94  & 0.063 & 0.005 & 0.706 & 0.15 \\
            2.06M  & 140 & 130 & 115 & 0.049 & 0.004 & 0.704 & 0.17 \\
            4.14M  & 170 & 160 & 154 & 0.041 & 0.003 & 0.696 & 0.17 \\
            6.14M  & 215 & 176 & 164 & 0.032 & 0.002 & 0.696 & 0.17 \\
            \hline \hline
        \end{tabular}
    \label{table:meshValidation}
    \end{center}
\end{table}
}

\end{appendices}

\FloatBarrier
\bibliography{vibrissa_literature}

\end{document}